\documentclass[11pt,twoside]{article}

\usepackage{asp2004}
\usepackage{epsf}
\usepackage{lscape}
\usepackage{graphicx}

\begin{document}
\title{3D-Kinematics of White Dwarfs from the SPY-Project}
\author{Roland Richter,$^1$ Uli Heber,$^1$ and Ralf Napiwotzki$^2$}
\affil{$^1$Dr.-Remeis-Sternwarte, Astronomisches Institut der Universit\"at Erlangen-N\"urnberg, Sternwartstra\ss e 7, 96049 Bamberg, Germany}
\affil{$^2$Centre for Astrophysics Research, STRI, University of Hertfordshire, College Lane, Hatfield AL10 9AB, UK}

\begin{abstract} 
We present a progress report on the kinematical analysis of the entire SPY
(ESO SN Ia Progenitor surveY; see Napiwotzki et al.\ 2001) sample of about one
thousand white dwarfs and hot subdwarfs. In a previous study (Pauli et
al.\ 2003, 2006) $398$ DA white dwarfs have been analysed already. Here we
extend the study to $634$ DA white dwarfs. We discuss kinematic criteria for a
distinction of thin disk, thick disk and halo populations. This is the largest
homogeneous sample of white dwarfs for which accurate 3D space motions have
been determined.  They have been derived from radial velocities, spectroscopic
distances and proper motions from catalogues. Galactic orbits and further
kinematic parameters were computed. Our kinematic criteria for assigning
population membership are deduced from a sample of F and G stars taken from
the literature for which chemical criteria can be used to distinguish between
thin disk, thick disk and halo members. The kinematic population
classification scheme is based on the position in the $VU$-velocity diagram,
the position in the eccentricity-$J_{\rm Z}$ diagram and the Galactic
orbit. We combine this with age estimates and find $12$ halo and $37$ thick
disk members amongst our DA white dwarfs. We were unable to determine the
population membership of only nine of them. The remaining members of the 
sample of 632 stars belong to the thin disk population.
\end{abstract}

\section*{Motivation}
White dwarfs are the evolutionary end-products of most stars. Therefore, a
large number of white dwarfs should be present in the Galaxy.  Determining the
contribution of white dwarfs to the total mass of the Galaxy could help to
solve one of the fundamental questions in modern astronomy: what is the nature
of dark matter? The fact that the rotation curves of many galaxies are not
Keplerian (Rubin, Thonnard, \& Ford 1978) invokes the existence of
additional dark matter distributed heavy-halo. It is estimated that for the
Milky Way only $10\%$ of the total mass are present in the form of stars, gas
and dust in the Galactic disk and halo (Alcock 2000). The role of white
dwarfs in the dark matter problem is still uncertain. An open issue is the
fraction of white dwarfs in the thick-disk and halo populations, as well as
their fraction of the total mass of the Galaxy. In this context, kinematic
studies have proved a useful tool in deciding on population membership of
white dwarfs.  The common problem of kinematic studies of white dwarfs is the
lack of radial velocity measurements. Especially deviating conclusions derived
from the white dwarfs of the Oppenheimer et al.\ (2001) sample demonstrate that
different assumptions on the values of the radial velocity 
$v_{\rm rad}$ can produce different
fractions of halo and thick disk stars and thus have a strong impact on the
determination of the white dwarf halo density (Reid, Sahu \& Hawley
2001). Therefore a sample of white dwarfs with known radial velocity
measurements is needed in order to obtain the full 3D kinematic
information. Pauli et al.\ (2006) presented a complete 3D kinematical study of
$398$ DA white dwarfs. Here we present a progress report on the analysis of
the entire SPY sample of more than one thousand white dwarfs and hot
subdwarfs. We present the results of a kinematic analysis of another $237$ DA
white dwarfs.

\section*{3D Kinematics and Population Classification}
The ESO Supernova Ia Progenitor surveY (SPY, see Napiwotzki et al.\ 2003) 
provided us with high resolution
spectra of about 1000 degenerate stars. The sharp absorption core of
the hydrogen lines allowed accurate radial velocities to be
measured. Typical errors are not greater than $\pm2$km/s. We added radial
velocity variable stars, for which the radial velocity curves had been
solved and the systemic velocity had been derived (Napiwotzki et al.\
2002, Karl et al.\ 2004 and Morales-Rueda et al.\ 2005). We also
included resolved wide binaries with no measured radial velocity
variability in two or more spectra ($\Delta v_R < 2$km/s). A
spectroscopic analysis, done by Voss and Koester (private
communications), yielded atmospheric parameters, which were used to
derive masses and gravitational redshifts. Proper motions were
extracted from catalogues through the VizieR Service (e.g.\ USNO--B,
UCAC2, etc.). We calculate individual errors of kinematic parameters
by means of Monte Carlo error propagation.

\begin{figure} [!ht]
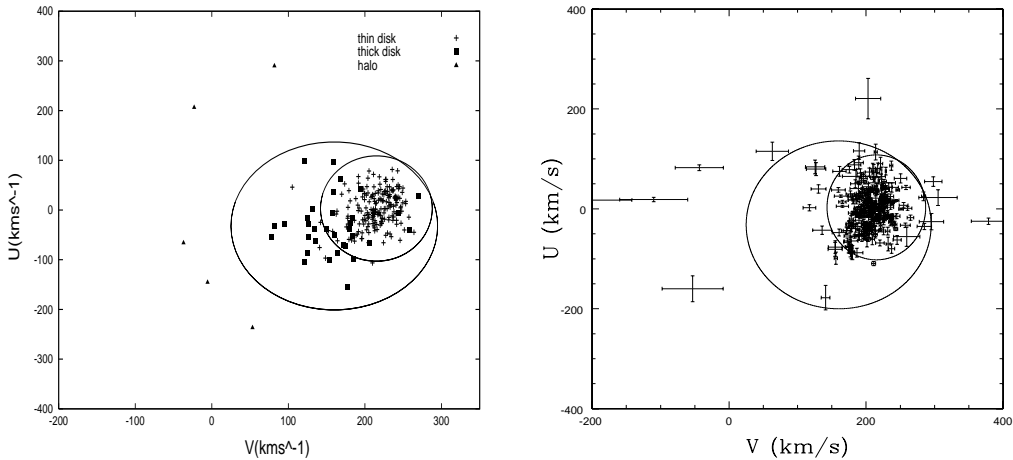

\includegraphics[keepaspectratio=false, width=0.47\textwidth, height=6.0cm]{newfig1a.epsi}\hfill
\includegraphics[keepaspectratio=flase, width=0.47\textwidth, height=6.0cm]{procVU.epsi}
\caption{$U$-$V$-velocity diagram for the main-sequence calibrators of Pauli et al.(2006, left) and of this paper (right); ellipses:
$3\sigma_{\rm thin}$-, $3\sigma_{\rm thick}$-contours.
}
\end{figure}

Pauli et al.\ (2003, 2006) presented a new population classification
scheme based on the $VU$-velocity diagram ($U$ and $V$ being the
velocity components in the Galactic plane), the eccentricity--angular
momentum ($e\/-J_{\rm Z}$) diagram and the Galactic orbit. For the
computation of orbits and kinematic parameters we used the code by
Odenkirchen \& Brosche (1992) based on the Galactic potential of Allen
\& Santill\'{a}n (1991).

Unlike for main-sequence stars the population membership of white dwarfs can not be determined from spectroscopically measured metalicities.
Therefore we have to rely on kinematic criteria, which have to be calibrated using a suitable calibration sample of main-sequence stars. In
our case this sample consists of $291$ F and G main-sequence stars from Edvardsson et al.\ (1993) and Fuhrmann (2004 and references cited
therein).

Halo and thick disk stars can be separated by means of their [Fe/H] abundances, they possess a higher [Mg/Fe] ratio than thin disk stars
(see Pauli et al., 2003, 2006 for details). In Figure 1 (left panel) $U$ is plotted versus $V$ for the main-sequence stars. For the
thin disk and the thick disk stars the mean values and standard deviations (3$\sigma$) of the two velocity components have been calculated.
The values for the thin disk are:
$\left<U_{\rm ms}\right>=3~\pm 35$ km/s$^{-1}$,
$\left<V_{\rm ms}\right>=215~\pm 24$ km/s$^{-1}$,
The corresponding values for the thick disk are:
$\left<U_{\rm ms}\right>=-32~\pm 56$ km/s$^{-1}$,
$\left<V_{\rm ms}\right>=160~\pm 45$ km/s$^{-1}$,
The values for the standard deviations are:
$\sigma_{U_{\rm ms}}=35$ km/s$^{-1}$,
$\sigma_{V_{\rm ms}}=24$ km/s$^{-1}$.
The corresponding values for the thick disk are:
$\sigma_{U_{\rm ms}}=56$ km/s$^{-1}$ and
$\sigma_{V_{\rm ms}}=45$ km/s$^{-1}$.
Indeed, nearly all thin disk stars stay inside the
$3\sigma_{\mathrm{thin}}$-limit and all halo stars lie outside
the $3\sigma_{\mathrm{thick}}$-limit, as can be seen from Figure 1.

\begin{figure}[!ht]
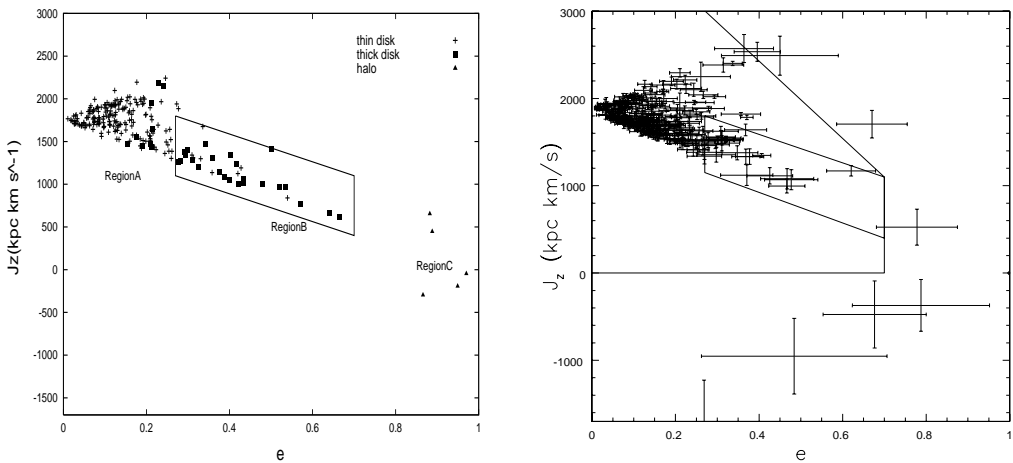

\includegraphics[keepaspectratio=false, width=0.47\textwidth, height=6.0cm]{newfig2a.epsi}\hfill
\includegraphics[keepaspectratio = flase, width=0.47\textwidth, height = 6.0cm]{proceJz.epsi}
\caption{$e-J_{\rm z}$-diagram for the main-sequence calibrators
of Pauli et al.\ (2006, left), Region A refers to the thin disk area,
Region B to the thick disk and Region C to the halo area on the graph;
the new data of this paper (right) shows also stars above Region B of
the calibration sample, which belong either to the thick disk (to the
left of the line) or to the halo (to the right of the line). We also
found white dwarfs with retrograde orbits (below the horizontal line).
}
\end{figure}

In the $e\/-J_{\rm Z}$-diagram three regions (A, B or C) can be
defined (see Figure~2, left panel) which host thin disk, thick disk
and halo stars, respectively. The Galactic orbits of thin disk, thick
disk stars and halo stars differ in a characteristic way allowing
another classification criterion to be defined (see Figure~3).

When plotted in a $\rho-Z$ diagram ($\rho=\sqrt{(X^2+Y^2)}$ with $X$,
$Y$, $Z$ being the rectangular Galactic coordinates), the populations
can be distinguished by their orbits' shape. Thin disk stars reach
only small altitudes above the galactic plane ($<600$pc, Figure 3,
left hand panel). Thick disk stars can climb up to $2.5$kpc (Figure 3,
middle panel) and finally halo stars reach even higher altitudes above
the plane or have a highly chaotic orbit (Figure 3, right hand panel).

Halo candidates are all white dwarfs that are either situated outside
the $3\sigma$-limit of the thick disk in the $VU$-velocity diagram or
that lie in Region C in the $e$-$J_{\rm Z}$ diagram and have halo type
orbits.  For the DA white dwarfs the $VU$-velocity diagram and the
$e$-$J_{\rm Z}$ are shown in the right hand panels of Figures 1 and 2
respectively.  In total $12$ white dwarfs fulfil the classification
criteria for halo stars and are therefore assigned to the
halo. Including some white dwarfs on retrograde orbits characterised
by a negative value of $V$ and $J_{\rm Z}$.

Thick disk white dwarfs lie either outside the $3\sigma$-limit of the
thin disk in the $VU$-velocity diagram or lie in Region B in the
$e-J_{\rm Z}$ diagram and have thick disk type Galactic
orbits. Thirty-seven of them are classified as thick disk
members. All, but nine, of the remaining are assumed to belong to the
thin disk. Those nine white dwarfs could not clearly be classified as a
member of a certain population, thus leaving us with $12$ halo, $37$
thick disk out of the $632$ SPY DA white dwarfs.

\begin{figure}[!ht]
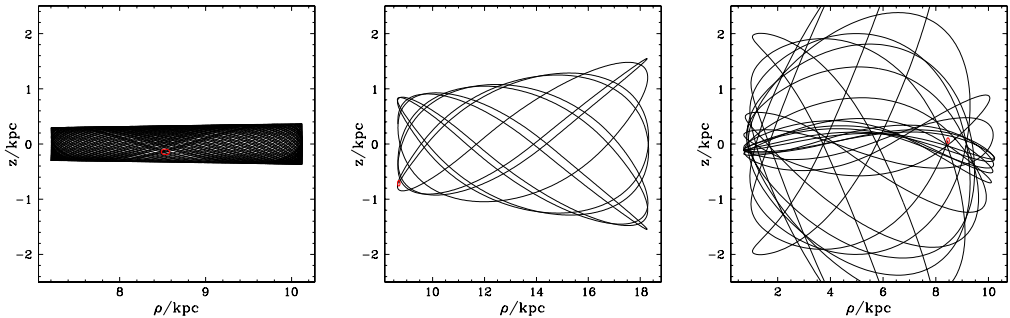

\centering
\includegraphics[width=0.31\textwidth]{newfig3a.epsi}\hfill
\includegraphics[width=0.31\textwidth]{newfig3b.epsi}\hfill
\includegraphics[width=0.31\textwidth]{newfig3c.epsi}
\caption{Typical orbits shown in a $\rho-Z$ diagram (with $\rho=\sqrt{(X^2+Y^2)}$).
Plotted are a thin disk (WD0017+061), a thick disk (WD0158-227) and a halo star (WD1314-153) (from left to right; please note the growing
scale of the $\rho$-axis). All orbits were calculated numerically with 500,000 year steps for 2Gy into the future.
}
\end{figure}

\begin{figure} [!ht]
\centering
\includegraphics[width=0.47\textwidth]{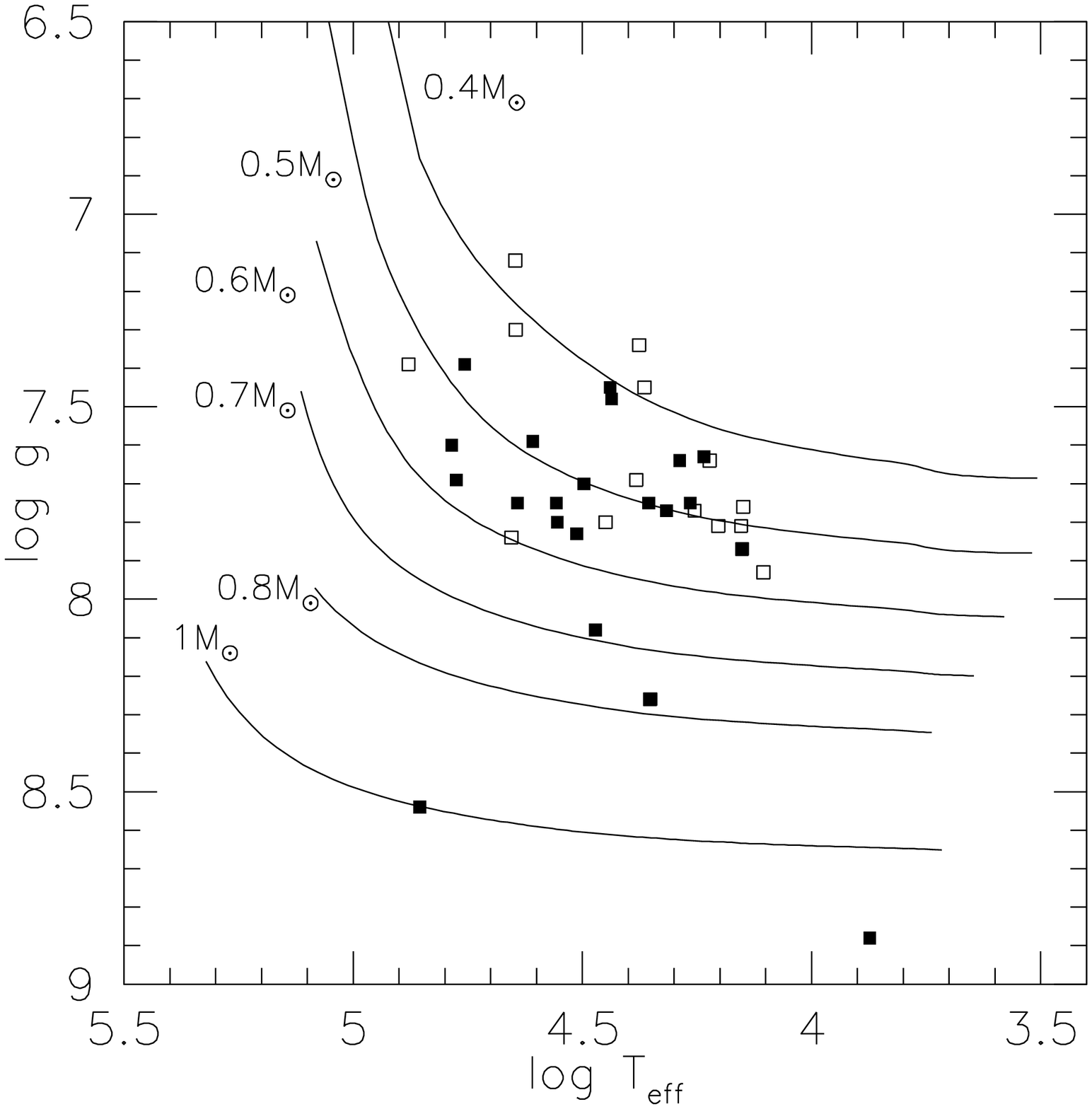}\hfill
\includegraphics[width=0.47\textwidth]{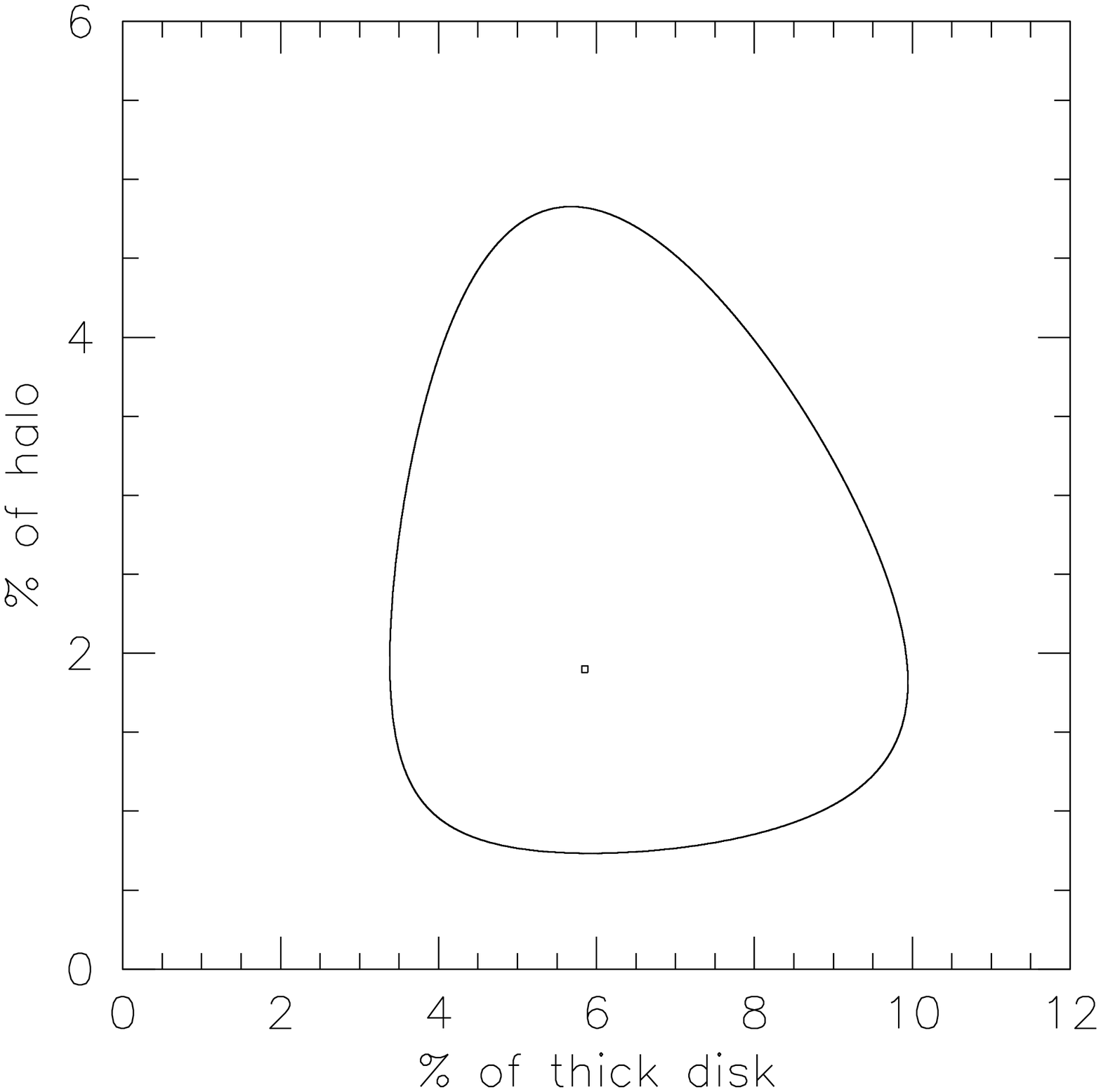}
\caption{left hand side: cooling sequences (Wood, 1995) and the values of the thick disk (filled squares) and halo (open squares) white 
dwarfs of this work. And on the right hand side: statistical results, with 3$\sigma$-contour.}
\end{figure}

\section*{The Age Test}
Ages are another criterion for population membership. Halo and thick
disk white dwarfs should be old stars. Since the cooling ages of the
of most white dwarfs in the SPY sample are small ($< 10^9$yr) they
have evolved from long-lived, i.e.\ low mass stars and hence must
themselves be of low mass.

We used the cooling tracks form Wood (1995) and Driebe et al.\ (1998) (see Figure 4, left hand side) to determine the masses and the cooling
times of our white dwarfs and the initial-to-final-mass
relation of Weidemann (2000) to get the mass of our progenitor stars.

All but one halo star and $31$ of the $39$ thick disk white dwarfs have
sufficiently low masses and are therefore old. However, our age estimates
could be misleading if the stars had evolved through close binary interaction
phases or may even result from a merger event. On the other hand the
kinematics of a thick disk or halo white dwarf could be mimiced by a thin disk
star ejected from the galactic plane (runaway star). The halo star 
WD\,0239+109 is most
likely such a runaway star from the thin disk, as it is only 2.2 Gyr
 old and is
not massive enough ($0.6M_{\odot}$) to be of a merger origin. Only one of
the thick disk stars has a high enough mass to be a merger ($1.0M_{\odot}$),
two live within a binary system. Hence we issue a birth certificate of the
thick disk population only to these $31$ white dwarfs. The others could be
runaway stars from the thin disk or merger, as their kinematic classification
is quite assured. This leaves us with a fraction of $2.3\%$ halo and $6.2\%$
thick disk white dwarfs. Figure 4 (right hand side) shows our final result
within the statistical 3$\sigma$-border.

\section*{Discussion}
The classification scheme developed in Pauli et al.\ (2003) has been used to
kinematically analyse a sample of more than $600$ DA white dwarfs from the SPY
project. Combining the three kinematic criteria, position in the $VU$-diagram,
position in the $e-J_{\rm Z}$ diagram and Galactic orbit, with age estimates
we have found twelve halo and $37$ thick disk members.

When we do the statistics for our local, white dwarf, population memberships
we find a 3$\sigma$-area as shown in Figure 4 (right hand side).  There's
still room for $3.5\%$ to $10.5\%$ of the local, hot white dwarfs to to belong
to the thick disk. And between $1.0\%$ and $5.3\%$ can belong to the halo
population and contributes a big amount to the baryonic dark matter of the
milky way.

\acknowledgements
Roland Richter thanks the RAS and the Astronomische Gesellschaft for financially supporting his participation in this workshop.

\end{document}